\documentclass[prl,aps,twocolumn]{revtex4} 
\usepackage{graphicx} 
\def\strutdepth{\dp\strutbox}
\def\nw#1{\strut\vadjust{\kern-\strutdepth\vtop to0pt{\vss\hbox to\hsize
{\hskip\hsize\hskip5pt$\leftarrow$\hss\strut}}}{\em #1}}
\begin{document} 
\title{Thick films coating a plate withdrawn from a bath} 

\author{J.H. Snoeijer$^1$, J. Ziegler$^1$, B. Andreotti$^2$,
        M. Fermigier$^2$, J. Eggers$^1$}
\affiliation{
School of Mathematics, University of Bristol, University Walk, 
Bristol BS8 1TW, UK$^1$ \\
PMMH, UMR 7636 CNRS-ESPCI-P6-P7, 10 rue Vauquelin, 
75231 Paris Cedex 05, France$^2$
              } 

\date{\today} 
 
\begin{abstract}
We consider the deposition of a film of viscous liquid on a flat plate
being withdrawn from a bath, experimentally and theoretically. For any 
plate speed $U$, there is a range of ``thick'' film solutions
whose thickness scales like $U^{1/2}$ for small $U$. These solutions
are realized for a partially wetting liquid, while for a perfectly wetting
liquid the classical Landau-Levich-Derjaguin (LLD) film is observed, whose
thickness scales like $U^{2/3}$. The thick film is distinguished from
the LLD film by a dip in its spatial profile at the transition to the 
bath. We calculate the phase diagram for the existence of stationary 
film solutions as well as the film profiles, and find excellent agreement 
with experiment. 
\end{abstract} 
 
\maketitle 

\begin{figure}[t!]
\includegraphics{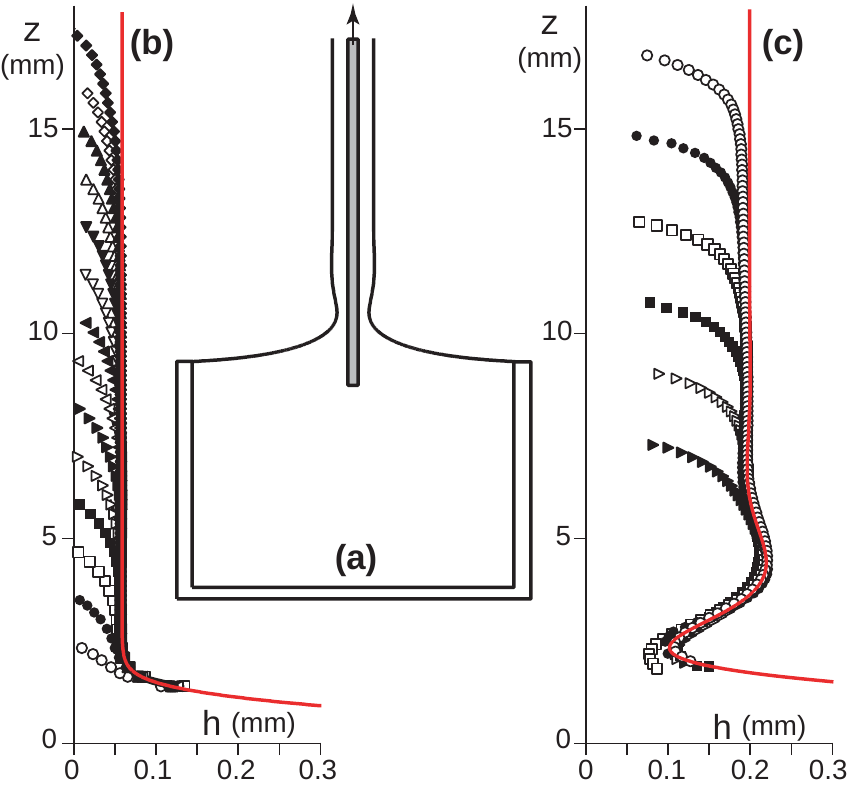}
\caption{ Fluid films being deposited on a plate withdrawn from
a bath of viscous liquid; symbols are experimental measurements 
of the film profile at successive times. 
{\bf (a)} Schematic of the experiment. 
{\bf (b)} The fluid wets the plate and deposits a LLD film, 
${\rm Ca}=9.27\times10^{-3}$; the solid line is the classical LLD solution,
which predicts $h_f=57\mu m$.
{\bf (c)} The plate is treated such that the fluid wets partially,
and deposits a thick film, ${\rm Ca}=8.05\times10^{-3}$;
the solid line is the result of our theory with the film thickness 
fitted to the experimental result of $h_f=198\mu m$. 
\label{fig.film} 
               }
\end{figure}

\begin{figure}[t!]
\includegraphics{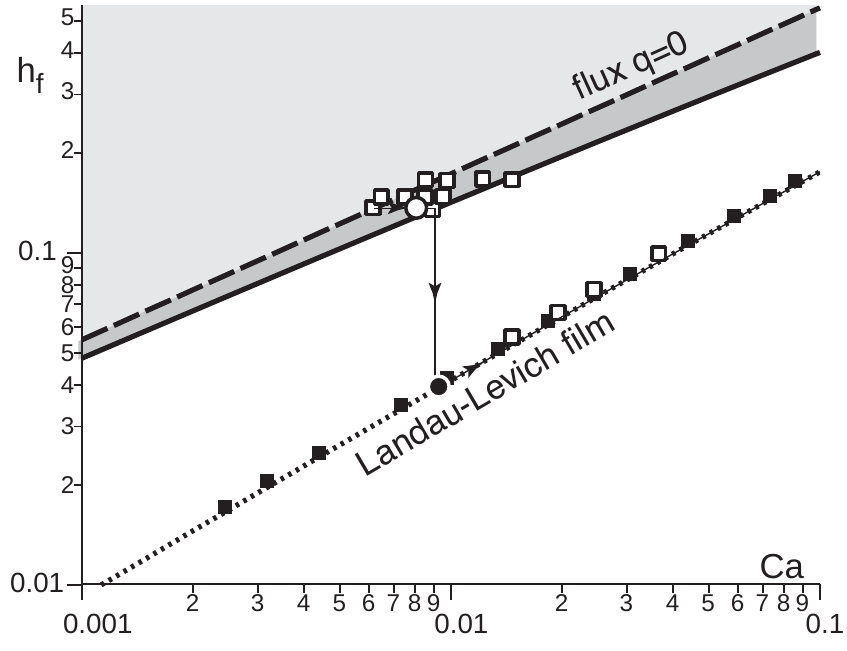}
\caption{Phase diagram of stationary solutions in the parameter 
space $(h_f,{\rm Ca})$. The dimple solutions exist in the light grey region, 
but only those in the dark region actually entrain liquid from the bath. 
The dotted line shows the Landau-Levich solution. The open symbols
are for a partial wetting situation, the full symbols for complete wetting;
error bars are below the size of the symbols. 
The two round symbols correspond to the two measurements shown in 
Fig.~\ref{fig.film}. The arrows show the experimental path as the
speed is slowly increased starting from the dashed line. 
       } 
\label{fig.hCa} 
\end{figure}

When a solid object is pulled out of a liquid reservoir, a thin layer 
of liquid is entrained by viscous drag (Fig.~\ref{fig.film}). This 
principle is used widely in coating technology, where it is known as 
dip coating, because it is one of the simplest ways to deposit a thin 
film of liquid on a substrate \cite{WeRu04}. According to the pioneering 
work of Landau \& Levich~\cite{LL42} and Derjaguin~\cite{D43} (LLD), 
a film of unique thickness $h_{LLD}$ is selected by the speed of 
withdrawal $U$. The LLD solution has remained the basis of coating theory
for more than 60 years, having been generalized to include the effects
of inertia \cite{RQ98a,JAM05}, deposition on curved substrates \cite{Q99},
and non-Newtonian fluids \cite{RQ98b}. 
The relative size of viscous drag and capillary retention in the film
is measured by the capillary number ${\rm Ca}=U\eta/\gamma$, where $\eta$ is 
the viscosity and $\gamma$ the surface tension. At the foot of the film,
LLD introduced the requirement that the film be matched smoothly to a 
static capillary meniscus (see Fig.~\ref{fig.film}), whose size is controlled
by the capillary length $\ell_c=\sqrt{\gamma/\rho g}$. In the limit of
small ${\rm Ca}$, this matching yields 
$h_{LLD}=0.946\ell_c {\rm Ca}^{2/3}$, which gives the small-${\rm Ca}$
behavior of the LLD line in the phase diagram, cf. Fig.~\ref{fig.hCa}. 

In this Letter, we show that at a given speed there exists another,
thicker film in addition to the LLD film (cf. Figs.\ref{fig.film}
and \ref{fig.hCa}). These new solutions {\it do not} match smoothly to
the bath, but exhibit a bump at the foot of the film 
(cf. Fig. \ref{fig.film} (c)).
As we will explain in more detail below, experimentally the thick 
solutions are most easily produced for partially wetting fluids just
above the critical plate speed ${\rm Ca}_{f}$ at which a film begins to be
deposited \cite{SDAF06}. Our theoretical analysis however is for the
long-time limit only, in which the plate is covered completely by a film, so
any solution is characterized by a pair $(h_f,{\rm Ca})$, as shown in 
Fig.\ref{fig.hCa}. 

A striking feature of the new solutions is that they cover a continuous 
range of thickness $h_f$ (Fig.~\ref{fig.hCa}, grey area). 
Each of these solutions entrain a different amount of liquid flux $q$, 
representing the volume of liquid withdrawn from the bath per unit 
time and unit plate length. Clearly, this flux is limited by the flow 
through the narrow dimple. Below we compute the maximum possible flux 
through the dimple and show how this provides a lower bound on 
$h_f$ (Fig.~\ref{fig.hCa}). For larger $h_f$, draining due to gravity 
becomes increasingly important, so at the upper bound 
(Fig.~\ref{fig.hCa}, dashed line) the flux vanishes. Above the dashed
line, no solutions are possible, unless the film is alimented from above.

In our theoretical analysis we rescale all lengths by the capillary 
length $\ell_c$. The interface profile is analyzed using the 
lubrication equation~\cite{footcurv} for free surface flows \cite{ODB97}: 
\begin{equation}\label{eq.lubrication}
\kappa'  - 1 + \frac{3}{h^2} \left({\rm Ca} - \frac{q}{h} \right)=0.
\end{equation}
Here $h(z)$ is the interface thickness, $\kappa$ the interface 
curvature and $q$ the flow rate. The three terms correspond 
to the capillary pressure, gravity, and viscous effects, respectively.
For steady solutions, considered here, $q$ is constant along the plate. 
Film solutions become asymptotically flat, ($\kappa=0$), so that 
the flux can be related to the film thickness $h_f$ as
\begin{equation}
q =h_f \left({\rm Ca} - \frac{h_f^2}{3} \right). 
\label{qh}
\end{equation}

\begin{figure}[t!]
\includegraphics{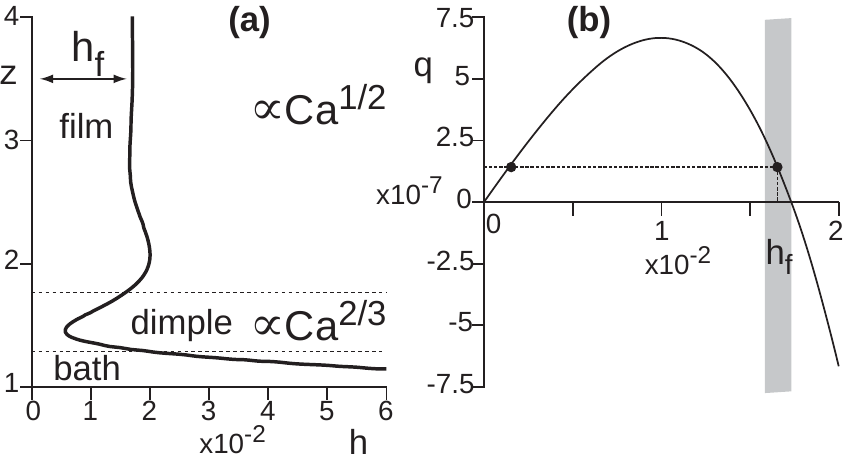}
\caption{{\bf (a)} The three asymptotic regions of the dimple solutions,
and the scaling of the respective film thickness with ${\rm Ca}$. 
{\bf (b)} The flux $q$ as function of $h_f$ at a fixed ${\rm Ca} = 10^{-4}$, 
cf. (\ref{qh}). The shaded region shows the range of thick solutions
at that value of ${\rm Ca}$. The horizontal line illustrates that the
flux through the thick film is the same as through the thin dimple region.
 } 
\label{fig.match} 
\end{figure}

Crucially, $q$ exhibits a non-monotonic dependence on $h_f$, owing to 
gravitational draining (Fig.~\ref{fig.match}b). For small $h_f$ the 
fluid velocity is uniform across the layer and equals the plate 
velocity, yielding a (dimensionless) flux $h_f{\rm Ca}$. For larger 
thickness gravity induces draining inside the film, which 
can even reverse the direction of liquid transport: $q<0$ corresponds to 
transport from the film into the reservoir. As a consequence, the same 
flow rate $q$ can be supported by two very different thicknesses of the 
film. This is why, physically, the thin 'dimple' region can be matched 
to the much thicker film (Fig.~\ref{fig.match}).

The interface structure in the dimple region results from a balance 
between viscosity and surface tension. To describe the profile near the 
dimple position $z_d$, we therefore look for a similarity 
solution of (\ref{eq.lubrication})
\begin{equation}
h={\rm Ca}^{2/3} H\left(\eta\right), \quad
\eta = (z-z_d)/{\rm Ca}^{1/3}. 
\label{ansatz_LLD}
\end{equation}
Following LLD, this solution has the additional property that the 
{\it curvature} $h''$ remains finite in the limit of small ${\rm Ca}$,
which we are analyzing. This ensures that (\ref{ansatz_LLD}) can
be matched to the capillary meniscus of the bath, which has a constant 
curvature of $\sqrt{2}$ \cite{LL84a}, so the boundary condition 
in the limit $\eta\rightarrow -\infty$ is $H_{-\infty}''=\sqrt{2}$. 
Defining $Q=q {\rm Ca}^{-5/3}$, one obtains the equation for the dimple 
\begin{equation}\label{eq.LL}
H''' + \frac{3}{H^2}\left(1 - \frac{Q}{H} \right)=0.
\end{equation}
So far (\ref{eq.LL}) is the same as for the LLD solution, but the 
boundary condition toward the flat film region will be less 
restrictive, so we are able to produce additional solutions. 
In that region, the curvature becomes small, giving the condition 
$H_{\infty}''=0$. Below we will justify this argument in more detail.

\begin{figure}[t!]
\includegraphics{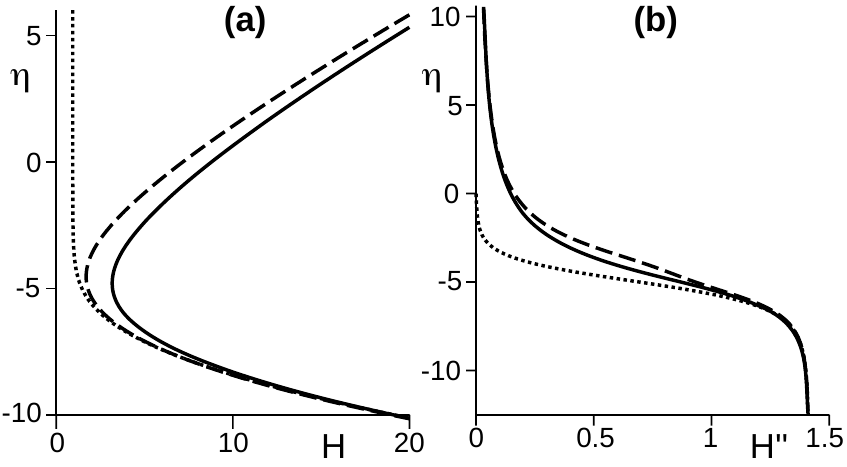}
\caption{Various solutions of (\ref{eq.LL}): 
{\bf (a)} Dimple profiles with $Q=0$ (solid), 
$Q_{\rm max}=1.376\cdots$ (dashed), and the Landau-Levich film 
solution for which $Q\approx 0.946$ (dotted). 
{\bf (b)} All curves have $H''_{-\infty}=\sqrt{2}$ to match the curvature 
of the reservoir, and $H''_{\infty}=0$ to match to the film.} 
\label{fig.dimple}
\end{figure}

Figure~\ref{fig.dimple} shows that there is a one-parameter family 
of solutions $H(\eta)$, each corresponding to a different value 
of $Q$, and forming a dimple that matches the thick film to the 
static bath. The case $Q=0$ can be solved analytically \cite{DW97}, 
while the others have been obtained numerically. We identified a 
maximum value $Q_{\rm max} \approx 1.376$, above which the boundary 
conditions can no longer be satisfied. Note that the LLD film is a 
particular solution of (\ref{eq.LL}), for which $H$ approaches a constant 
value rather than shooting back upwards (dotted line). 

The final step is to relate the flux through the dimple to the film thickness 
$h_f$ at a large distance from the reservoir. Just before $q(h_f)$ changes 
sign, it becomes small enough to match the flux in the dimple 
(cf. Fig.~\ref{fig.match}).
For a small enough capillary number, (\ref{qh}) can be solved to give
\begin{equation}\label{eq.range}
h_f \simeq \left(3{\rm Ca} \right)^{1/2} - \frac{1}{2} Q {\rm Ca}^{2/3},
\end{equation}
where $Q \in \left[0,Q_{\rm max} \right]$. The dimple with $Q_{\rm max}$ 
corresponds to the lowest possible thickness (lower bound of grey
region in Fig.~\ref{fig.hCa}), while the case $Q=0$ marks the upper 
boundary for $h_f$ with nonnegative flux. It is worth noting that 
$Q=0$ is only the upper bound in the case of the dip-coating geometry. 
If an injection of liquid is added at the top of the plate, one could 
match much thicker films to the bath. The structure of the films will 
become rather different when the downward flow is much larger than 
the plate velocity, for which the analysis crosses over to flow down a 
wall at rest \cite{WJ83}. 

The similarity solution of (\ref{eq.lubrication}) corresponding to
the film region just above the dimple is 
\begin{equation}
h={\rm Ca}^{1/2} \widetilde{H}\left(\frac{z-z_d}{{\rm Ca}^{1/6}}\right). 
\label{thick_ss}
\end{equation}
In (\ref{thick_ss}), the $h$-scale ${\rm Ca}^{1/2}$ is dictated by the
film thickness (\ref{eq.range}) for small ${\rm Ca}$. Thus the typical 
curvature $h'' \propto {\rm Ca}^{1/6}$ vanishes in the limit 
${\rm Ca}\ll 1$, which establishes the boundary condition 
$H_{\infty}''=0$ quoted above. The slopes in the dimple and film regions,
as given by (\ref{ansatz_LLD}) and (\ref{thick_ss}), respectively, 
both scale as ${\rm Ca}^{1/3}$ and can thus be matched. To establish 
our central result (\ref{eq.range}), a more detailed analysis of
the film solutions (\ref{thick_ss}) is not necessary.

Note that the final approach onto the flat film involves (stationary) 
capillary waves on the film surface that are exponentially damped 
(see Fig.~\ref{fig.match} {\bf (a)}). From the point of view of the 
lubrication equation (\ref{eq.lubrication}), these waves provide an 
additional degree of freedom that is necessary to achieve different values of 
flux \cite{H01,JAM05,DJZ07}. Linearizing around the flat film, one 
finds that such waves can in principle exist when $h_f > {\rm Ca}^{1/2}$. 
As can be seen from (\ref{eq.range}), however, this condition is 
not sufficient to explain the new films. Similar dimpled solutions have
also been found for Marangoni-driven flows \cite{ME05}. 

\begin{figure}[t!]
\includegraphics{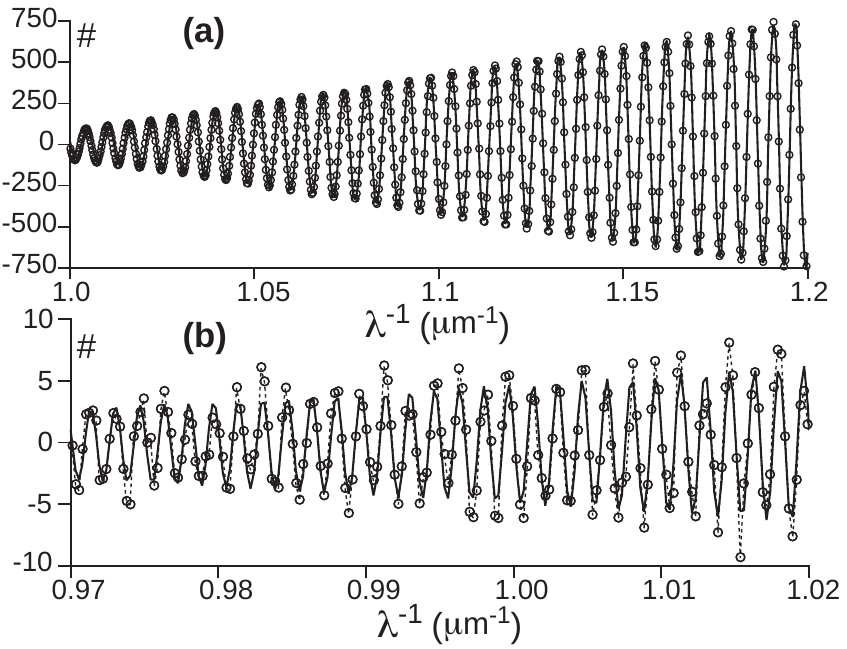}
\caption{Measurement of the film thickness by spectroscopy. The free 
charge carriers produced by photon absorption are accumulated during 
$0.4$~s, counted and displayed as a function of the inverse wavelength 
$\lambda^{-1}$. This spectrum is averaged over $100$ realizations 
and corrected from the effect of CCD heterogeneities, calibrated
previously. It is then split into the sum of a smooth spectrum, 
characteristic of the light source, and an oscillating spectrum, shown 
above (symbols). It is finally fitted by a sine modulated in 
amplitude by a Gaussian envelope (solid line). 
{\bf (a)} LLD film at $Ca=9.75\times10^{-2}$.  The best fit gives 
$h_f=61.24$~nm with (formally) a statistical accuracy of $0.005$~nm. However,
the reproducibility of two complete experimental runs is only to within 
$0.1$~nm. 
{\bf (b)} Thick film at $Ca=8.53\times10^{-2}$. The best fit gives 
$h_f=214.68$~nm with (formally) a statistical accuracy of $0.05$~nm and a 
reproducibility to within $1$~nm. Thick films are at the limit of 
resolution, hence the interference pattern is only visible in the 
near-infrared range of wavelengths.} 
\label{fig.spect}
\end{figure}

{\it Experimental methods~--}~In our experiment, a silicon wafer is 
withdrawn vertically from a bath of silicone oil 
(viscosity $\eta=4.95 \,{\rm Pa\!\cdot\! s}$, surface tension 
$\gamma=0.0203 \,{\rm N \!\cdot\! m^{-1}}$, 
density $\rho=970 \,{\rm kg\!\cdot\! m^{-3}}$, molecular size $70$~nm) 
by a step-motor. The silicon wafer is totally wetted by the silicone oil. 
When coated with the fluorinated surfactant FC725, partial wetting 
conditions are obtained. Depending on the protocol, the contact angle 
lies between $48^\circ$ and $55^\circ$, with an hysteresis of $5^\circ$. 
The film thickness $h_f$ is measured by spectrometry. A reflection probe 
made of a tight bundle of six illumination fibers around one read fiber 
is placed at $5$~mm from the plate. It is connected to a Tungsten Halogen 
light source and to a diffraction-grating spectrometer resolving 
visible and near-infrared wavelengths. The relative resolution, limited 
by the entrance slit and the size of the photosensitive elements, is 
around $0.25$~nm. A 3648-element linear CCD-array detector measures 
the intensity as a function of the wavelength $\lambda$, averaged over 
$4s$. Defects of the CCD are calibrated in the absence of film and 
corrected to improve the signal-to-noise ratio. Once a film is present, 
the spectrum contains oscillations of the form $\cos(4 \pi n h_f/\lambda)$, 
where $n=1.4034$ is the refractive index of silicone oil. By fitting the 
spectrum, the thickness $h_f$ can be measured to within $0.2\%$ up to a 
thickness of $240~\mu$m (cf. Fig.~\ref{fig.spect}).

The spatial profile of the entrained film is measured by placing a $200~\mu$m 
wire  at a distance $d=14 \pm 1$~mm  from the silicon wafer. The mirror 
image of the wire (reflected in the silicon plate) is distorted by 
refraction through the liquid/vapor interface of local slope $h'$, 
making the system equivalent to a wire placed at a distance 
$d$ {\it behind} a prism of (small) angle $2h'$. At small angles, 
the rays are thus deflected by an angle $2 (n-1) h'$ in the direction 
of steepest slope, independent of the incident angle. Hence the mirror image 
of the wire is vertically displaced by a distance $2 (n-1) d h'$. 
The wire is imaged with a $2048$ x $2048$ CCD camera, fitted with a $60$~mm 
macro lens. The position of the wire image is determined by image 
inter-correlation, achieving sub-pixel resolution. 

We have calibrated the relation between the displacement and the local 
interface slope by two independent methods. First, we have replaced the 
oil film by optical glass prisms ($n=1.52$) of angle 
$1^\circ$, $2^\circ$, $3^\circ$ and $4^\circ$. We obtain a resolution of 
$950 \pm 30~$pixels per unit slope ($16.5~{\rm pixels}/\circ$), once the 
index difference between oil and glass is taken into account. Second, 
using the step-motor, we have determined an image resolution of 
$11.7~\mu$m/pixels, which leads to $970 \pm 70~$pixels per unit slope. 
To summarize, the displacement of the image of the wire gives the local 
interface slope from which we reconstruct the film profile, by integration. 
The constant of integration is fixed in the flat region of the film, 
using the value of $h_f$ measured by spectrometry. There are two sources 
of error in the final reconstruction: an absolute uncertainty of $3\%$ in 
the slope, and the error from the spatial integration of the random 
noise present for each point.

Finally, we return to the question of how a particular type of film
solution is realized experimentally. As shown in \cite{SDAF06}, the
front of a liquid film of partially wetting fluid possesses a characteristic
speed of recession ${\rm Ca}_f$ relative to the substrate, which is 
essentially set by the contact angle. Since the flux $q$ through the 
contact line is zero, according to (\ref{qh}) this translates into a 
characteristic {\it film thickness} $h_f = (3{\rm Ca}_f)^{1/2}$. Thus if the
plate speed ${\rm Ca}$ is marginally above ${\rm Ca}_f$, the contact 
line moves up the plate very slowly, and the flux through the liquid 
film is close to zero: one has prepared the state $(h_f,h_f^2/3)$
at the upper end of the thick film solutions. 

If the plate speed is now slowly increased, as we did in an experimental 
run shown in Fig.~\ref{fig.hCa} by the arrows, one moves horizontally 
into the grey region, since $h_f$ is fixed
by the motion of the contact line. As the end of the grey region is 
reached, the solution falls off the region of allowed solutions, 
and a LLD solution is realized instead (cf. Fig.~\ref{fig.hCa}). 
During this transition, the dimple detaches from the bath, 
and moves up the plate, leaving behind the LLD film. In effect,
this is the situation described in our earlier paper \cite{SDAF06}.
Note that we were not able to take a measurement right above the 
transition, since the LLD film takes too long to develop. 
Of course the above does not imply that thick films can only be produced 
in partial wetting situations. The most promising experimental setup
to produce a thick film is that of viscous flow in the interior of 
a cylinder, which can be rotated in both senses. This offers 
considerably more flexibility in preparing thicker films. 

In conclusion, we have calculated a phase diagram for film coating from
a bath. At a given speed, two different types of film solutions 
are possible. The novel, thick films described here are easily realized
experimentally, and experiment agrees extremely well with theoretical
prediction. 

\begin{acknowledgments}
We gratefully acknowledge Howard Stone and Wiebke Drenckhan for 
clarifying discussions and Giles Delon for sharing his experimental 
expertise. J.H.S. acknowledges support from a Marie Curie European 
Action FP6 (Grant No. MEIF-CT-2006-025104).
\end{acknowledgments}


\begin{thebibliography}{16}
\bibitem[{foo()}]{footcurv}
\bibinfo{note}{At small ${\rm Ca}$ viscous effects become negligable whenever
  $h'\mbox{\ \raisebox{-.9ex}{$\stackrel{\textstyle >}{\sim}$}\ } 1$, so that
  the equation remains valid in the bath.}
\end{thebibliography}
\end{document}